\documentclass[12pt]{article}
\usepackage{graphicx}

\begin{document}
\title{\Huge\bf Kinematics of the internal space associated with the TEGR}
\author{{{L.R.A. Belo}$^{\,a}$\thanks{E-mail: leandrobelo@fis.unb.br}, {V.C. de Andrade}$^{\,a}$\thanks{E-mail: andrade@fis.unb.br}, {E.P. Spaniol}$^{\,a}$\thanks{E-mail: spaniol@fis.unb.br}, and {J.A. de Deus}$^{\,a}$\thanks{E-mail: julianoalves@fis.unb.br}}\\ \\
\it $^a$ \small \it Physics Institute, Brasilia University\\ \small \it 70.917-910, Brasilia, Federal District, Brazil}

 \date{}
\maketitle
\begin{abstract}
In the context of Teleparallel Equivalent of General Relativity - TEGR - 
we have obtained, through the second kind gauge transformations, the 
most fundamental transformations, namely, the first kind ones. We show that considering the possibility of decomposing the components of the tetrad field as a trivial part plus some potential
besides the usual translational and Lorentz potentials, there is
also the possibility that the symmetry group of internal space be a
generalization of the Poincar\'e group. Still in the analysis of
transformations in the internal space, we saw that for the case in
which the decomposition of the tetrad field components includes only
a trivial part plus a translational potential, we are able to recover just the
translational group. Moreover, the internal space of gravity must be
the very physical space - on a local scale - in such a way that the
gauge symmetry group is the kinematical group of the physical space itself. This group was obtained, and seems to generalize the Poincar\'e group.

\textbf{keywords:} Teleparallel Equivalent of General Relativity, First kind gauge transformations, Kinematics.

\end{abstract}

\newpage

\section{Introduction}
\indent Gauge theories of gravitation are an attempt to describe such interaction using techniques well established in the description of the other fundamental interactions. Moreover, the study of symmetries and conservation laws in physical space is a powerful tool for understanding the fundamentals of mechanics. The idea of extrapolating and using the same kind of apparatus in an internal space opens a window of possibilities; this was exactly what allowed us to understand the real essence of the "charge": quantities conserved under the action of a particular group of symmetry in internal space \cite{aldrovandi3}.

The existence of an internal space, in turn, is associated with extra degrees of freedom which are evident in the hamiltonian formulation of the theory. These theories are said to be constrained, since its phase space has decreased \cite{dirac}. Although in the study of constrained hamiltonian systems the appearance of extra degrees of freedom (internal space) is a consequence of the size of the configuration space, without any physical notion of real essence of this space, we know they are degrees of freedom associated with the source field; a consequence of the invariance of the total Lagrangian under transformations in this field.

The equivalence principle states that the equations of special relativity must be recovered in a locally inertial coordinate system in which gravitational effects are absent. Thus, based on this principle, it would be natural to expect that gravitation had a local Poincar\'e symmetry, and it was possible to describe it as a genuine gauge theory for this group. In fact, this is possible \cite{blagojevic}. However there are evidences, both theoretical and experimental/observational, that from the most fundamental point of view (beyond the standard model) the Lorentz symmetry is broken \cite{Kostelecky,Carroll,Colladay,Belich,Songaila,Moffat,Coleman}. This would eliminate the Poincar\'e group as local symmetry group of gravity, leaving room only for the translational sector or another more general.

The TEGR is a particular case of a more general theory that has a set of three free parameters, the Teleparallel Gravity, which is fully equivalent to General Relativity \cite{hayashi,aldrovandi,andrade,arcos,aldrovandi2}. In order of better understanding its internal structure, one should apply the tools mentioned in the paragraphs above. As will be seen, we show that starting from the TEGR Lagrangian presented in the literature, we conclude that the internal space associated with the TEGR seems to be invariant under the action of a group which generalizes the Poincar\'e group.

Notation: the Greek alphabet is used to represent physical space indices ($\alpha,\beta,\chi$,... = 0,1,2,3), the first half of the Latin alphabet $(a,b,c.. = 0,1,2,3)$ represents indices of the internal space, or gauge, and the second half of the latin alphabet $(i,j,k...)$ assumes the values of 1, 2 and 3 of the physical space.

\section{The TEGR and their First Class Constraints as Generators of the Gauge Transformations}
The Lagrangian density associated with TEGR has the form \footnote{The torsion is defined by $T^\rho{}_{\mu \nu} \;\equiv\Gamma^\rho{}_{\nu \mu} \;-\Gamma^\rho{}_{\mu \nu}$. The object $\Gamma^\rho{}_{\nu \mu}$ is the Weitzenbock connection defined by $\Gamma^\rho{}_{\nu \mu} \;\equiv h_{a}{}^{\rho}\partial_{\mu}h^{a}{}_{\nu}$.} \cite{andrade}:
\begin{equation}
{\cal L} = \frac{h}{2k^2} \left[\frac{1}{4} \; T^\rho{}_{\mu \nu} \; T_\rho{}^{\mu \nu} + \frac{1}{2} \; T^\rho{}_{\mu \nu} \; T^{\nu \mu}{}_\rho - T_{\rho \mu}{}^{\rho} \; T^{\nu \mu}{}_\nu \right] \; \label{1}
\end{equation}
with  $h=det(h^{a}{}_{\mu})$ and $k=\frac{8\pi G}{c^4}$. This expression can be rewritten in a more elegant form to get \cite{maluf}:
\begin{equation}
{\cal L}=\frac{h}{4k^2}\;S^{\rho \mu \nu }\;T_{\rho \mu \nu }\;, \label{2}
\end{equation}
where
\begin{equation}
S^{\rho \mu \nu }=-S^{\rho \nu \mu }\equiv {\frac{1}{2}}\left[ K^{\mu \nu \rho }-g^{\rho \nu }\;T^{\theta \mu }{}_{\theta }+g^{\rho \mu }\;T^{\theta \nu }{}_{\theta }\right] \label{3}
\end{equation}
and $K^{\mu \nu \rho }$ is the contortion tensor given by
\begin{equation}
K^{\mu \nu \rho}= \frac{1}{2} \, T^{\nu \mu \rho}+\frac{1}{2} \, T^{\rho \mu \nu}- \frac{1}{2}T^{\mu \nu \rho}. \label{4}
\end{equation}

The procedure of a Legendre transformation from the previous Lagrangian density is not sufficient to get its Hamiltonian version, since it is not possible to isolate all the terms of velocities depending on the momenta \cite{belo}. To circumvent this problem, we should simplify this Lagrangian density, making it linear before proceeding \cite{maluf2}; this "linearization" is done by introducing auxiliary fields $\phi_{abc}=-\phi_{acb}$ that will be related to the torsion tensor. The first order differential Lagrangian formulation in empty physical space reads
\begin{equation}
{\cal L}=kh\Lambda^{abc}\left(\phi_{abc}-2T_{abc}\right), \label{5}
\end{equation}
where $T_{abc}=h_{b}{}^{\mu}h_{c}{}^{\nu}T_{a\mu\nu}$, $\Lambda^{abc}$ is defined by
\begin{equation}
\Lambda^{abc}=\frac{1}{4}\left(\phi^{abc}+\phi^{bac}-\phi^{cab}\right)+\frac{1}{2}\left(\eta^{ac}\phi^{b}-\eta^{ab}\phi^{c}\right), \label{6}
\end{equation}
and $\phi_{b}=\phi^{a}{}_{ab}$. After a long development, the first class secondary constraints are found \footnote{These is the relevant constraints in the calculation of the gauge transformations.} 
\begin{equation}
\chi_{c}=h_{c}{}^{0}{\cal H}_{0}+h_{c}{}^{i}F_{i}, \label{7}
\end{equation}
where
\begin{eqnarray}
{\cal H}_{0}&=&-h_{a 0}\partial_{k}\Pi^{a k}-\frac{kh}{4g^{0 0}}(g_{i k}g_{j l}P^{i j}P^{k l}-\frac{1}{2}P^2) \nonumber \\
&+&kh(\frac{1}{4}g^{i m}g^{n j}T^{a}{}_{m n}T_{a i j}+\frac{1}{2}g^{n j}T^{i}{}_{m n}T^{m}{}_{i j}-g^{i k}T^{j}{}_{j i}T^{n}{}_{n k}) \label{8}
\end{eqnarray}
and
\begin{equation}
F_{i}=h_{a i}\partial_{k}\Pi^{a k}-\Pi^{a k}T_{aki}+\Gamma^{m}T_{0mi}+\Gamma^{lm}T_{lmi}+\frac{1}{2g^{0 0}}(g_{i k}g_{j l}P^{k l}-\frac{1}{2}P)\Gamma^{j}. \label{9}
\end{equation}
Furthermore, the objects were defined:
\begin{equation}
\Pi^{ai}=-4kh\Lambda^{a0i}, \label{10}
\end{equation}
\begin{eqnarray}
P^{ik}&=&\frac{1}{2kh}(h_{c}{}^{i}\Pi^{ck}+h_{c}{}^{k}\Pi^{ci})+g^{0m}(g^{kj}T^{i}{}_{mj}+g^{ij}T^{k}{}_{mj}-2g^{ik}T^{j}{}_{mj}) \nonumber \\
&+&(g^{km}g^{0i}+g^{im}g^{0k})T^{j}{}_{mj}, \label{11}
\end{eqnarray}
\begin{equation}
\Gamma^{ik}=\frac{1}{2}(h_{c}{}^{i}\Pi^{ck}-h_{c}{}^{k}\Pi^{ci})-kh[-g^{im}g^{kj}T^{0}{}_{mj}+(g^{im}g^{0k}-g^{km}g^{0i})T^{j}{}_{mj}] \label{12}
\end{equation}
and
\begin{equation}
\Gamma^{k}=\Pi^{0k}+2kh(g^{kj}g^{0i}T^{0}{}_{ij}-g^{0k}g^{0i}T^{j}{}_{ij}+g^{00}g^{ik}T^{j}{}_{ij}); \label{13}
\end{equation}
where $\Pi^{ai}$ are the momenta canonically conjugated to $h_{ai}$. Being the constraints first class we can use them to calculate the transformations generated by the constraints, which do not alter the physical state of the system (gauge transformations) \cite{dirac}:
\begin{eqnarray}
\delta h^{b}{}_{\rho}(x)&=&\int d^{3}x'
\left[\varepsilon_{1}^{a}(x')\{ h^{b}{}_{\rho}(x),\Phi_{a}(x)
\}+\varepsilon_{2}^{a}(x')\{ h^{b}{}_{\rho}(x),\chi_{a}(x) \} \right] \nonumber \\
&=&\int d^{3}x' \varepsilon_{1}^{a}(x') \left(\frac{\delta
h^{b}{}_{\rho}(x)}{\delta h^{c}{}_{\beta}(x')} \frac{\delta
\Phi_{a}(x)}{\delta \Pi_{c}{}^{\beta}(x')}-\frac{\delta
h^{b}{}_{\rho}(x)}{\delta \Pi_{c}{}^{\beta}(x')} \frac{\delta
\Phi_{a}(x)}{\delta h^{c}{}_{\beta}(x')} \right) \nonumber \\
&+&\int d^{3}x' \varepsilon_{2}^{a}(x') \left(\frac{\delta
h^{b}{}_{\rho}(x)}{\delta h^{c}{}_{\beta}(x')} \frac{\delta
\chi_{a}(x)}{\delta \Pi_{c}{}^{\beta}(x')}-\frac{\delta
h^{b}{}_{\rho}(x)}{\delta \Pi_{c}{}^{\beta}(x')} \frac{\delta
\chi_{a}(x)}{\delta h^{c}{}_{\beta}(x')} \right), \label{14}
\end{eqnarray}
that results in \cite{belo}
\begin{equation}
\delta h^{b}{}_{\rho}=\delta^{0}_{\rho}\varepsilon^{b}_{1}+\nabla_{\rho}\varepsilon_{2}^{b}, \label{15}
\end{equation}
with
\begin{equation}
\nabla_{\rho}\varepsilon_{2}^{b}\equiv\delta_{\rho}^{i}\partial_{i}\varepsilon_{2}^{b}+\omega^{b}{}_{a \rho} \varepsilon_{2}^{a} \label{16}
\end{equation}
and
\begin{eqnarray}
\omega^{b}{}_{a \rho}&\equiv&-\frac{1}{g^{00}}\delta_{\rho}^{i}h_{a}{}^{0}h^{b}{}_{i}g^{0\mu}T^{j}{}_{j\mu}+\frac{3}{2g^{00}}\delta^{i}_{\rho}g^{0 b}h_{a i}g^{0\mu}T^{j}{}_{j\mu}+\frac{1}{2g^{00}}\delta^{i}_{\rho}h^{b 0}h_{a i}g^{0\mu}T^{j}{}_{j\mu} \nonumber \\
&+&\frac{3}{2}\delta^{i}_{\rho}h^{b}{}_{\mu}h_{a}{}^{\nu}T^{\mu}{}_{i \nu}+\delta^{i}_{\rho}g^{0 b}g_{0 \mu}h_{a}{}^{\nu}T^{\mu}{}_{i \nu}-\frac{1}{2}\delta^{i}_{\rho}g_{i \mu}h^{b \nu}h_{a}{}^{\alpha}T^{\mu}{}_{\nu \alpha} \nonumber \\
&+&\frac{1}{2}\delta^{i}_{\rho}h_{a i}h^{b \mu}T^{0}{}_{0 \mu}
-\delta^{i}_{\rho}h^{b}{}_{i}h_{a}{}^{\mu}T^{0}{}_{0 \mu}+\frac{1}{2}\delta^{i}_{\rho}\delta^{b}_{a}T^{0}{}_{0 i} \label{3.20}
\end{eqnarray}
playing the role of covariant derivative and connection, respectively. We can go ahead and rewrite (\ref{15}) as follows:
\begin{equation}
\delta h^{b}{}_{\rho}=\delta^{0}_{\rho}\varepsilon^{b}_{1}+\delta_{\rho}^{i}\partial_{i}\varepsilon_{2}^{b}+\omega^{b}{}_{a \rho} \varepsilon_{2}^{a}. \label{18}
\end{equation}
Introducing now the following relation between the parameters $\varepsilon^{b}_{1}$ and $\varepsilon^{b}_{2}$
\begin{equation}
\varepsilon^{b}_{1}=\partial_{0}\varepsilon_{2}^{b}, \label{19}
\end{equation}
we have:
\begin{equation}
\delta h^{b}{}_{\rho}=\partial_{\rho}\varepsilon_{2}^{b}+\omega^{b}{}_{a \rho} \varepsilon_{2}^{a}. \label{20}
\end{equation}
The subindex 2 can now be ignored,
\begin{equation}
\delta h^{b}{}_{\rho}=\partial_{\rho}\varepsilon^{b}+\omega^{b}{}_{a \rho} \varepsilon^{a}\equiv \nabla^{'}_{\rho}\varepsilon^{b}. \label{21}
\end{equation}
The above transformations allow a direct analogy with the gauge transformations obtained in the Yang-Mills theory.

\section{Kinematics of Internal Space}
The transformations (\ref{21}) enable a fundamental analysis to understand the kinematics of the internal space (gauge symmetry group). We have
\begin{equation}
\delta h^{b}{}_{\rho}=\partial_{\rho}\varepsilon^{b}+\omega^{b}{}_{a \rho} \varepsilon^{a}, \label{22}
\end{equation}
and motivated by the soldering property \cite{kobayashi}, we can decompose $h^{b}{}_{\rho}$ as follows
\begin{equation}
h^{b}{}_{\rho}=\delta_{\rho}^{b}+A^{b}{}_{\rho}+A_{a}{}^{b}{}_{\rho}x^{a}+.... \label{23}
\end{equation}
Even if there is no gauge potential (gravity), the trivial part
$\delta_{\rho}^{b}$ remains "linking" the physical and internal
spaces. Although we have no idea of what would represent other
potentials, besides the already well known $A^{b}{}_{\rho}$ and
$A_{a}{}^{b}{}_{\rho}$ \footnote{Components of the gauge potential
associated with the translational and Lorentz sectors, which arises
as compensating fields that guarantee the invariance of Lagrangian
under first kind gauge transformations.}, we have kept open the
possibility that they exist. Armed with this decomposition, it is
possible to write
\begin{equation}
\delta h^{b}{}_{\rho}=h^{b'}{}_{\rho}-h^{b}{}_{\rho}=\delta_{\rho}^{b'}-\delta_{\rho}^{b}+\delta\left(A^{b}{}_{\rho}\right)+\delta\left(A_{a}{}^{b}{}_{\rho}x^{a}\right)+..., \label{24}
\end{equation}
or, using (\ref{22})
\begin{equation}
\frac{\partial}{\partial x^{\rho}}\left(x^{b'}-x^{b}\right)=\partial_{\rho}\varepsilon^{b}+\omega^{b}{}_{a \rho} \varepsilon^{a}-\delta A^{b}{}_{\rho}-\delta \left(A_{a}{}^{b}{}_{\rho}x^{a}\right)-... \label{25}
\end{equation}
From this point it is possible, by integrating in $x^{\rho}$, to get the following expression for the internal space transformations:
\begin{equation}
\delta x^{b}=\epsilon^{b}+C^{b}-F^{b}-\varpi^{b}{}_{a}x^{a}-... \label{26}
\end{equation}
with
\begin{eqnarray}
\epsilon^{b}&\equiv& \int \partial_{\rho}\left(\varepsilon^{b}\right)dx^{\rho}, \nonumber \\
C^{b}&\equiv& \int \omega^{b}{}_{a \rho}\varepsilon^{a}dx^{\rho}, \nonumber \\
F^{b}&\equiv& \int \delta A^{b}{}_{\rho} dx^{\rho}, \nonumber \\
\varpi^{b}{}_{a}x^{a} &\equiv& \int \delta \left(A_{a}{}^{b}{}_{\rho}x^{a}\right) dx^{\rho}. \label{27}
\end{eqnarray}
Or, rearranging terms:
\begin{equation}
\delta x^{b}=\epsilon'^{b}+\varpi'^{b}{}_{a}x^{a}+... \label{28}
\end{equation}
here $\epsilon'^{b}\equiv \epsilon^{b}+C^{b}-F^{b}$, and the negative sign of $\varpi^{b}{}_{a}$ was absorbed by $\varpi'^{b}{}_{a}$. In this case, assuming that the last integral in (\ref{27}) results in the usual Lorentz transformations, we clearly have a generalization of the Poincar\'e transformations
\begin{equation}
\delta x^{b}=\epsilon^{b}+\varpi^{b}{}_{a}x^{a}; \label{29}
\end{equation}
however, this generalization only exists for cases in which we include potentials beyond the usual $A^{b}{}_{\rho}$ and $A^{a b}{}_{\rho}$, which would imply the existence of other degrees of freedom associated with these potentials. For the case where the last integral in (\ref{27}) does not result in the Lorentz transformations, we have a generalization even if we have not included other potential. In addition, the transformations (\ref{28}) may vary both in form and scale, the first two integrals in (\ref{27}) depend of $\varepsilon\left(x^{\mu}\right)$, and therefore have different results for each of the possible indexes for this parameter, internal or physical.

From the expressions (\ref{27}) and (\ref{28}) it can be noted that for the case in which $h^{b}{}_{\rho}$ is decomposed only as $\delta^{b}_{\rho}+A^{b}{}_{\rho}$, all the "rotation" matrices $\varpi^{b}{}_{a}$ are zero, hence:
\begin{equation}
\delta x^{b}=\epsilon'^{b}, \label{30}
\end{equation}
which is exactly a translation in the internal space. This is in total agreement with the literature of gauge theories to the translation group \cite{aldrovandi}, in which the fundamental object is the potential $A^{b}{}_{\rho}$.

Regardless of what values are calculated in the integrations (\ref{27}) \footnote{The solutions depend on the specific form of the fields to be calculated.}, it seems reasonable to think that their results will take the form (\ref{28}), since through it we can get both Poincar\'e as translations, depending on how we make the decomposition of the tetrad field components $h^{b}{}_{\rho}$.

\section{Universality through a gauge approach}
From the viewpoint of gauge theories, the gauge fields "appear" as
compensating fields, that guarantee the invariance of the Lagrangian
under the action of a transformation group acting on the source
field (first kind gauge transformations) \cite{aldrovandi3}.
Thinking this way, the analysis of a Lagrangian without sources -
anywhere - is merely an academic exercise, not having much
physical support; the only way to "generate" the gauge field is
through the action of the symmetry group of the internal space
(variation of the source field). For the case of gravitation,
however, the previous statement is not necessarily true, since the
free Lagrangian leads to the field equations that contain a term of
"vacuum source" \footnote{Gravitational Maxwell's equations enable a proper analysis of this issue \cite{spaniol}.}: in other words,
gravitation is nonlinear. When trying to do a mental exercise and
view the generation of gauge fields in the case of gravitation, we
face the following problem. Let us consider an arbitrary source of
gravity defined in an internal space. Acting with the symmetry group
of this space there emerges a distortion in a compensatory way
(tension, making an analogy with a fabric) in physical space, which
manifests itself through the gravitational field. This gravitational
field, in turn, having a not null energy should also generate an
additional distortion in space (second order effect), but, being
defined in physical space, how could it restart the cycle? It is
necessary that the gravitational field interferes locally in the
internal space for it generates, in turn, a new perturbation in the
physical space (second order field), and this process repeats itself indefinitely. The Figure 1 illustrates some of what was said.

\begin{figure}[h]
	\centering
		\includegraphics[height=0.3\textheight]{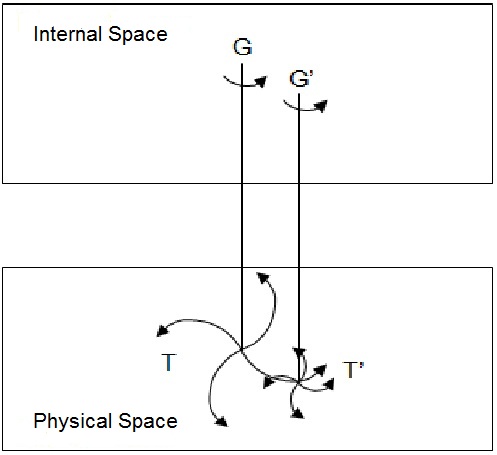}
	\caption{Distortions of the first and second order $T$ and $T'$, representing the field strength generated by the acting of the groups $G$ and $G'$.}
	\label{fig:Figura1}
\end{figure}

Observing the first and second kind gauge transformations it is possible to infer how the process occurs. We have, after a gauge transformation in the internal space,

\begin{equation}
x^{b}\rightarrow x^{b'}=x^{b}+\delta x^{b}=x^{b}+\epsilon'^{b}+\varpi'^{b}{}_{a}x^{a}+..., \label{31}
\end{equation}
and consequently the physical space responds with the emergence of
$h^{b}{}_{\rho}$.  This, in turn, varies with $\varepsilon^{b}$
according to
\begin{equation}
\delta h^{b}{}_{\rho}=\nabla^{'}_{\rho}\varepsilon^{b}=\partial_{\rho}\varepsilon^{b}+\omega^{b}{}_{a \rho} \varepsilon^{a}. \label{32}
\end{equation}
The connection $\omega^{b}{}_{a \rho}$ can be divided into two parts;
the first one leads algebra indices into algebra indices again, as
in the Yang-Mills theory. The second, in turn, leads algebra indices
into physical space indices:
\begin{equation}
\omega^{b}{}_{a \rho}\varepsilon^{a}=\left(..._{\rho}\right)\delta_{a}^{b}\varepsilon^{a}+\left[\left(...^{b}{}_{\rho}\right)h_{a}{}^{0}+\left(...^{b}\right)h_{a\rho}+\left(...^{b}{}_{\rho \nu}\right)h_{a}{}^{\nu}\right]\varepsilon^{a}, \label{33}
\end{equation}
or,
\begin{equation}
\omega^{b}{}_{a \rho}\varepsilon^{a}=\left(..._{\rho}\right)\varepsilon^{b}+\left[\left(...^{b}{}_{\rho}\right)\varepsilon^{0}+\left(...^{b}\right)\varepsilon_{\rho}+\left(...^{b}{}_{\rho \nu}\right)\varepsilon^{\nu}\right]. \label{34}
\end{equation}
The first part is analogous to the Yang-Mills connection. The second
part, however, allows that $\varepsilon$, now with physical space index, calibrates objects defined in this space. Note that this is not only a consequence of the locality of gauge transformations; the
effect of the first kind transformations is just "to force"
the emergence of the field. However, it does not ensure that this
field generates new transformations; these arise as consequence of
the parameters $\epsilon'^{b}$ having the possibility to be written as
being $\epsilon'^{\mu}$. In addition, as already mentioned, the form of transformations (\ref{31}) is not fixed, it varies with $x^{\mu}$; the first two integrals
in (\ref{27}) depend on $\varepsilon$, and therefore they have different
results for each of the possible indices for this parameter,
internal or physical.

Another key issue is that for the case of gauge theories such as
electromagnetism, for example, the first kind transformations are
represented by the action of the symmetry group in a given source field. For the case of gravitation, however, the action of the group takes place directly on the base that defines the internal space
$x^{b}$, not on a source field defined in this space \footnote{It is
clear that for cases in which there is a source field, besides the
gravitational field, the group will also act in this field.}. This
allows us to conjecture that the source of gravity must be
associated to the existence of internal space itself, and not to the sources defined therein, as with other interactions.

The equivalence principle, or equivalently the universality of free
fall, leaves a tiny gap to the issue of symmetry of internal space.
In fact, being the gravitation universal, its symmetry group of the
internal space should act on any field that has non-zero energy,
thus ensuring its coupling to this field. However, the very usual
concept of gauge symmetry rejects this idea, i.e. the gauge group acts
only on fields with specific properties \footnote{Such as the gauge
currents $J_{a}{}^{\mu}$.} and therefore defined in a special space,
called internal space. One way to circumvent this issue is to assume
that the internal space of gravity is the very physical space in a
"local scale", in such way that thinking in a symmetry group for
this space is now the same as thinking in a kinematics to the
physical space. This kinematics is given by the set of
transformations (\ref{28}), which can generalize the usual transformations of special relativity. For obvious reasons, this alternative is valid only if the gravitation is in fact universal.

The Figure 2 illustrates how would be the Fiber bundle of the fundamental interactions in the scenario described throughout this section.

\begin{figure}[h]
	\centering
		\includegraphics[height=0.4\textheight]{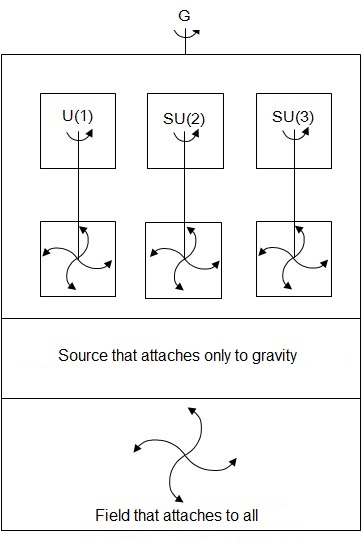}
	\caption{Physical space on local scale.}
	\label{fig:Figura2}
\end{figure}

Note that the scenario above is quite different from a mere extrapolation of gravity for the context of gauge theories. Such extrapolation would imply a scenario as illustrated by the Figure 3.

\begin{figure}[h]
	\centering
		\includegraphics[height=0.4\textheight]{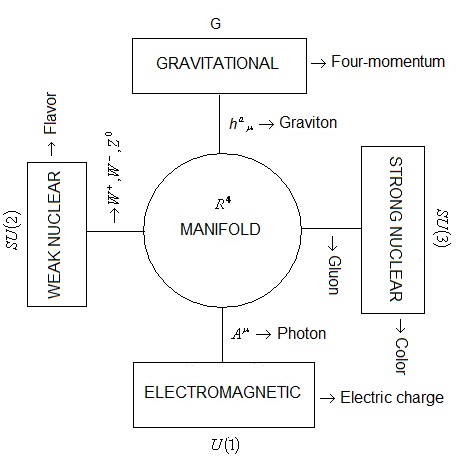}
	\caption{Representation of a Fiber bundle structure for the four fundamental interactions.}
	\label{fig:Figura3}
\end{figure}

However, as we have seen, there are sufficient reasons to disregard this second scenario.

\section{Final remarks}

Through the second kind gauge transformations, it was
possible to get the most fundamental transformations, i.e., the first
kind ones. We show that with the possibility of decomposing the tetrad
field components as a trivial part plus some potential besides the
usual Lorentz and translational potentials, there is still the
possibility that the symmetry group of the internal space be a
generalization of the Poincar\'e group. Still in the analysis of
transformations in the internal space, we saw that for the case in
which the decomposition of the tetrad field components mentioned
above includes only a trivial part plus the translational potential,
we recover the group of translations. This is in complete agreement
with the literature.

The \textit{spin} connection present in (\ref{32}), and rewritten in (\ref{33})
and (\ref{34}), allows that internal space indices be taken as physical
space indices, i.e., the parameters of gauge transformation can also
be defined on the base manifold. This unique peculiarity of gravitation makes impossible that the parameters defined in (\ref{27}) be
interpreted as phases of wave functions defined in internal space,
being this peculiarity related to universality of gravitation, i.e.,
the universality allows that objects be defined both in physical
space or in the internal space. In short, given a source field
$\psi$ \footnote {From the nonlinearity of field equations, we know
that $\psi$ can be associated with the gravitational field itself.
However, as already said, it does not mean that there are other
sources.}, $h^{a}{}_{\mu}$ can be any $h^{a}{}_{\mu}+\delta
h^{a}{}_{\mu}$, with $\delta h^{a}{}_{\mu}$  given by (\ref{22}). When
choosing a specific calibration, through the introduction of new
second class constraints, $\psi$ and $h^{a}{}_{\mu}$ are also
specified, i.e. all degrees of freedom have now been transferred to
the physical space. This specification, however, was not made at the
expense of choosing a particular phase of a wave function defined in
internal space, but choosing parameters, even without a precise
physical interpretation, which can be defined in both spaces. Is
also worth mentioning that being gravitation universal, its internal space must be the actual physical space (local scale), and
therefore its symmetry group of internal space is the kinematical
group of physical space itself. This group was obtained, and is
represented by the set of transformations (\ref{28}).

Finally, we are investigating the possible relationship between the clear non-compact appearance of the symmetry group of the internal space \footnote{This feature can be seen in (\ref{28}).} and the existence of gravitational singularities.

\begin{figure}[h]
	\centering
		\includegraphics[height=0.3\textheight]{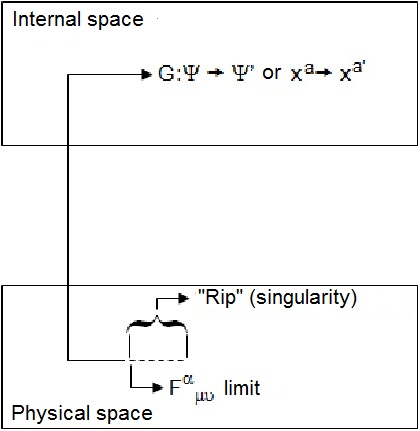}
	\caption{Representation of a rip in physical space, generated from the action of non-compact sector of the symmetry group of the internal space.}
	\label{fig:Figura4}
\end{figure}

From the astrophysicist viewpoint a singularity occurs when a gravitational energy source is confined to a sufficiently small region of space. From a more fundamental point of view, singularities like Schwarzschild $r_{S}$ exist independently of the value of the mass, being necessary only that the mass be contained in a region smaller than $r_ {S}$. The "non-compactness" allows the internal space to be "bigger" than the very physical space (in a local scale), thus, by making a first kind gauge transformation it is possible that the physical space, being locally attached to the internal space, has its structure "stretched" beyond the supportable. As a result, the fabric of the physical space would be "ripped" thus creating a singularity. The Figure 4 illustrates what was said. A complete analysis of this issue will be made elsewhere.

\vskip .8cm
\centerline{\bf \Large Acknowledgements\\}
\vskip .5cm \noindent The authors thank CAPES and CNPq (Brazilian agencies) for the financial support.\\

\end{document}